\documentclass[10pt,a4paper]{article}
\usepackage[utf8]{inputenc}
\usepackage{amsmath}
\usepackage{amsfonts}
\usepackage{amssymb}

\usepackage[electronic]{ifsym}
\usepackage{indentfirst,latexsym}
\usepackage{bm}
\usepackage{bbm}
\usepackage{mathrsfs}
\usepackage{amsmath,amsfonts,amsthm,amscd,amssymb}
\usepackage{pifont}
\usepackage{orcidlink}
\usepackage{url}

\usepackage{graphicx}
\usepackage[left=2cm,right=2cm,top=2cm,bottom=2cm]{geometry}

\usepackage{algorithm}         
\usepackage{algorithmicx}
\usepackage{algpseudocode}
\usepackage{float}

\newcommand{\Fig}{\textbf{Figure}~}
\newcommand{\Tab}{\textbf{Table}~}
\newcommand{\Thm}{\textbf{Theorem}~}

\newcommand{\ProcName}[1]{\textsc{#1}}

\algnewcommand\algorithmicswitch{\textbf{switch}}
\algnewcommand\algorithmiccase{\textbf{case}}
\algnewcommand\algorithmicdefault{\textbf{default}}

\algdef{SE}[SWITCH]{Switch}{EndSwitch}[1]{\algorithmicswitch\ #1\ \algorithmicdo}{\algorithmicend\ \algorithmicswitch}%
\algdef{SE}[CASE]{Case}{EndCase}[1]{\algorithmiccase\ #1}{\algorithmicend\ \algorithmiccase}%
\algdef{SE}[DEFAULT]{Default}{EndDefault}[1]{\algorithmicdefault\ #1}{\algorithmicend\ \algorithmicdefault}%

\makeatletter
\renewcommand{\ALG@name}{Algorithm}

\renewcommand{\bf}[1]{\mathbf{#1}}

\newcommand{\mfloor}[1]{ \left\lfloor {#1} \right\rfloor }
\newcommand{\mceil}[1]{ \left\lceil {#1} \right\rceil }

\newcommand{\mat}[1]{\bm{#1}}

\newcommand{\set}[1]{\left\{ #1 \right\}}
\newcommand{\seq}[1]{\left\langle #1 \right\rangle}
\newcommand{\abs}[1]{\left| #1 \right|}

\newcommand{\trsp}[1]{{#1}^\textsf{T}}

\newcommand{\inv}[1]{#1^{-1}}

\newcommand{\ES}[3]{\mathbb{#1}^{{#2}\times {#3}}}     

\DeclareMathOperator{\GCD}{GCD}

\newcommand{\scrd}[2]{{#1}_{\mathrm{#2}}}
\newcommand{\scrud}[3]{{#1}^{\mathrm{#2}}_{\mathrm{#3}}}

\newtheorem{thm}{\Thm }
\newtheorem{cor}[thm]{Corollary}

\newtheorem{res}[thm]{Result}

\author{Hong-Yan Zhang$^{a,b}$\orcidlink{0000-0002-4400-9133}, Haoting Liu$^{b,a}$, Zhi-Qiang Feng$^a$, Ci-Fei Dong$^a$,\\
 Rui-Jia Lin$^c$, Yu Zhou$^a$ and Fu-Yun Li$^a$\\
\begin{tabular}{l}
$^a$\small{\textit{School of Information Science and Technology, Hainan Normal University, Haikou 571158, P. R. China}}\\
$^b$\small{\textit{School of Automation, University of Science and Technology Beijing, Beijing 100083, P. R. China}}\\
$^c$\small{\textit{Information Network and Data Center, Hainan Normal University, Haikou 571158, P. R. China}}
\end{tabular}
}
\title{Pulling Back Theorem  for Generalizing the Diagonal Averaging Principle in Symplectic Geometry Mode Decomposition and \\ Singular Spectrum Analysis\thanks{Corresponding author: Hong-Yan Zhang, e-mail: hongyan@hainnu.edu.cn; ORCID: 0000-0002-4400-9133}}

\begin{document}
\maketitle

\begin{abstract}
The \textit{symplectic geometry mode decomposition} (SGMD) is a powerful method for analyzing time sequences. The SGMD is based on the upper conversion via embedding and down conversion via \textit{diagonal averaging principle} (DAP) inherited from the \textit{singular spectrum analysis} (SSA). However, there are two defects in the DAP: it just hold for the time delay $\tau=1$ in the trajectory matrix and it fails for the time sequence of type-1 with the form $X=\set{x[n]}^N_{n=1}$. In order to overcome these disadvantages, the inverse step for embedding is explored with binary Diophantine equation in number theory. The contributions of this work lie in three aspects: firstly, the pulling back theorem is proposed and proved, which state the general formula for converting the component of trajectory matrix to the component of time sequence for the general representation of time sequence and for any time delay $\tau\ge 1$; secondly a unified framework for decomposing both the deterministic and random time sequences into multiple modes is presented and explained; finally, the guidance of configuring the time delay is suggested, namely the time delay should be selected in a limited range via balancing the efficiency of matrix computation and accuracy of state estimation. It could be expected that the pulling back theorem will help the researchers and engineers to deepen the understanding of the theory and extend the applications of the SGMD and SSA in analyzing time sequences. 
\end{abstract}

\section{Introduction}

The \textit{symplectic geometry mode decomposition} (SGMD)  is originally proposed by Pan et al. in 2019 \cite{Pan2019sgmd} for decomposing  \textit{time sequence} (TS) which is also named with \textit{time series}. The SGMD is a development of symplectic geometry spectral analysis \cite{Xie2005SpGeoSpec,Xie2014symplectic} and Takens' delay-time embedding theorem \cite{Takens1981} for analyzing nonlinear signals or time sequences.  
The SGMD employs the symplectic geometry similarity transform \cite{Salam2008} to determine the eigenvalues of the Hamilton matrix and build a series of single components, known as the \textit{symplectic geometry components} (SGCs). These components together maintain the intrinsic characteristics of the TS involved while minimizing modal confusion. Moreover, the building process effectively eliminates noise \cite{Guo2022dynamic}, demonstrating strong decomposition performance and robustness against noise. 

In recent five years, there are active researches about the SGMD, such as 
Jin et al. \cite{Jin2019}, Zhang \cite{ZhangGY2022Esgmd}, Guo et al. \cite{GuoJC2022, GuoJC2023}, Chen et al. \cite{ChenYJ2023}, 
 Liu et al. \cite{Liu2024SpSparse}, Hao et al. \cite{Hao2024}, Zhan et al \cite{ZhanPM2024}, and Xin et al. \cite{XinG2025Csgmd}. 
Although the researches which  demonstrate the merits of the SGMD method are still increasing, there is \textit{no doubt about the key principle  behind the SGMD}.  Actually, there are two defects about the formula of diagonal averaging for converting the \textit{component of trajectory matrix} (CRM) to the 
\textit{component of time sequence} (CTS) in the original and various version of improved SGMD:
\begin{itemize}
\item it just holds for the time delay $\tau = 1$ in the step of constructing the trajectory matrix for embedding;
\item it works for the type-1 time sequence denoted by $X=\set{x[n]}^N_{n=1}$ but  fails for the type-0 time sequence denoted by $X=\set{x[n]}^{N-1}_{n=0}$  due to different structures of the trajectory matrix.
\end{itemize}  

In this work, our contributions to the SGMD lies in three aspects in the sense of theory and practice: 
\begin{itemize}
\item[i)] the pulling back theorem is proposed and proved, which state the correct form of the formula for converting the CRM to the CTS for the time sequence $X$ of type-$s$ denoted by $X=\set{x[n]}^{N-1+s}_{n=s}$ for $s\in \set{0, 1}$ and time delay $\tau\in \mathbb{N}$; 
\item[ii)] a unified framework for decomposing both the deterministic and random time sequences into multiple modes is presented and explained;
\item[iii)] the guidance of configuring the time delay is suggested --- it should be selected in a limited range with a trade off between the efficiency of matrix computation and accuracy of state estimation.
\end{itemize} 
The contents of this paper are organized as follows: Section \ref{sect-preliminaries} copes with the preliminaries about the notations and binary Diophantine equation; Section \ref{sect-embedding-ts} deals with the general embedding process for generating the trajectory matrix for the time sequence; Section \ref{sect-pullback} covers the pulling back theorem, which is the key issue of this work; Section \ref{sect-discussion} is the discussion of the SGMD and the pulling back theorem; finally, the conclusions are summarized in Section \ref{sect-conclusions}.

For the convenience of reading, some nomenclatures and notations are give in \Tab \ref{tab-nomenclatures}.
\begin{table*}[htbp] 
\caption{Nomenclatures and notations} \label{tab-nomenclatures}
\begin{tabular}{p{4cm}p{12cm}}
\hline
Notation & Interpretation \\
\hline
TS  & time sequence, also named with time series \\
CTM & Component of Trajectory Matrix \\
SGMD & Symplectic Geometry Mode Decomposition \\
SSA  & Singular Spectrum Analysis \\
SGC  & Symplectic Geometric Component \\
ISGC & Initial Symplectic Geometry Component \\
$\mathbb{Z}$ & set of integers, $\mathbb{Z} = \set{0, \pm 1, \pm 2, \cdots}$ \\
$\mathbb{Z}^+$ & set of non-negative integers $\mathbb{Z}^+ = \set{0, 1, 2, \cdots}$ \\
$\mathbb{N}$ & set of positive integers $\mathbb{N} = \set{1, 2, 3, \cdots}$ \\
$\Omega, \Omega(\alpha_1, \alpha_2, \beta_1, \beta_2)$ & rectangular domain such that $\Omega = \set{(x,y)\in \mathbb{Z}^2: \alpha_1\le x\le \alpha_2, \beta_1\le y \le \beta_2}$ \\
$\mceil{x}$ & ceiling of $x\in \mathbb{R}$, the minimal integer $ n \in  \mathbb{Z}$ such that $ n\ge x$ \\
$\mfloor{x}$ & floor of $x\in \mathbb{R}$, the maximal integer $ n\in \mathbb{Z}$ such that $ n\le x$  \\
$X = \set{x[n]}^{N-1+s}_{n=s}$ & time sequence of type-$s$ with length $N\in \mathbb{N}$ for $s\in \set{0,1}$ \\
$X^{(k)} = \set{x^{(k)}[n]}^{N-1+s}_{n=s}$ & the $k$-th CTS of type-$s$ with length $N\in \mathbb{N}$ for $s\in \set{0,1}$ such that $\displaystyle X = \sum^r_{k=1} X^{(k)}$\\
$\ES{R}{d}{m}$ & set of $d\times m$ matrices, immersion space of trajectory matrices \\
$(\mathscr{X}, \scrd{n}{dof})$ & manifold of time sequence with dimension $\scrd{n}{dof}$\\
$\Phi_{s,\tau}: \mathscr{X}\to \ES{R}{d}{m}$ & embedding mapping for up conversion \\
$\inv{\Phi}_{s,\tau}: \ES{R}{d}{m}\to \mathscr{X}$ & pulling back mapping for down conversion, inverse of embedding mapping \\
$\mat{A} = (A^i_j)_{d\times m}\in \ES{R}{d}{m}$ &  $d\times m$ matrix, where $A^i_j$ is the entry located in the $i$-th row and $j$-th column \\
$\mat{M}=\Phi_{s,\tau}(X)$ & trajectory matrix such that $ M^{i+s}_{j+s}= (x[i \tau +j +s])$ for $s\in \set{0,1}$\\
$\mat{Z}_k=(Z^i_j(k))_{d\times m}$ & the $k$-th CTM of the trajectory matrix $\mat{M}$ \\
$(\mat{Z}_1, \cdots, \mat{Z}_r)$ & group of $r$ CTMs such that $\displaystyle \mathcal{D}(\mat{M}) = \sum^r_{k=1}\mat{Z}_k$\\
$(\hat{\mat{Z}}_1, \cdots, \hat{\mat{Z}}_{\hat{r}})$ & group of  $\hat{r}$ CTMs after denoising\\
$\mathcal{D}: \ES{R}{d}{m}\to \ES{R}{d}{m}$ & matrix decomposition, $\displaystyle \mathcal{D}(\mat{M}) = \sum^r_{k=1} \mat{Z}_k$  \\
$\mathcal{F}: \ES{R}{d}{m}\to \ES{R}{d}{m}$ & denoising operation in the immersion space, $\displaystyle \sum^{\hat{r}}_{k=1}\hat{\mat{Z}}_k= \mathcal{F}\left(\sum^r_{i=1}\mat{Z}_i \right)$ \\
$\Psi: \mathscr{X}\to \mathscr{X}$ & SGMD mapping such that $\Psi = \inv{\Phi}_{s,\tau}\circ \mathcal{D}\circ \Phi_{s,\tau}$ without denoising or
$\Psi = \inv{\Phi}_{s,\tau}\circ \mathcal{F}\circ \mathcal{D}\circ \Phi_{s,\tau}$ with denoising \\
\hline
\end{tabular}
\end{table*}

\section{Preliminaries} \label{sect-preliminaries}

\subsection{Notations}

The time sequence $X$ of length $N\in \set{N}$ can be denoted by $x[n]$ for simplicity. However, there are two types of concrete representations for different computer programming languages:
\begin{itemize}
\item type-1 for the Fortran/MATLAB/Octave/...
      \begin{align*}
       X = \set{x[n]: 1\le n\le N}
         = \seq{x[1], x[2], \cdots, x[N]}
      \end{align*}
\item type-0 for the C/C++/Python/Java/Rust/...
      \begin{align*}
      X = \set{x[n]: 0\le n\le N-1} 
        = \seq{x[0], x[1], \cdots, x[N-1]}
      \end{align*}
\end{itemize}
It is trivial to find that the unified formula for these two types can be expressed by
\begin{equation}
X=\set{x[n]: s\le n\le N-1+s}=\set{x[n]}^{N-1+s}_{n=s}
\end{equation}
for $s\in \set{0,1}$. 
In signal processing, the time sequence $X=
\set{x[n]}^{N-1+s}_{n=s}$ is usually denoted by $x[n]$ for simplicity. An alternative notation for time sequence in mathematics and physics is $x_n$. Moreover, the time sequences are also named by discrete time signals or time series.

The manifold of time sequences $X$ ot type $s\in \set{0, 1}$  is denoted by $(\mathscr{X}, \scrd{n}{dof})$, where $\scrd{n}{dof}$ is the dimension of the space $\mathscr{X}$. Note that $\scrd{n}{dof}$ is the freedom of degeree for the dynamic system $x(t)$ for $t\ge t_0$ such that 
\begin{equation}
x[n] = x(t_0 + n/f), \quad n\in \mathbb{Z}^+
\end{equation}
where $f$ is the sampling frequency and $t_0$ is the initial time. Usually, the $x[n]$ can be regarded as the sum of the ideal signal $\scrd{x}{ideal}[n]$ and the  
noise $\scrd{x}{noise}[n]$, namely
\begin{equation}
x[n] = \scrd{x}{ideal}[n] + \scrd{x}{noise}[n], \quad n\in \mathbb{Z}^+
\end{equation}

\subsection{Binary Diophantine Equation}

\subsubsection{General Case}

\begin{figure}[h]
\centering
\includegraphics[width=0.5\textwidth]{ 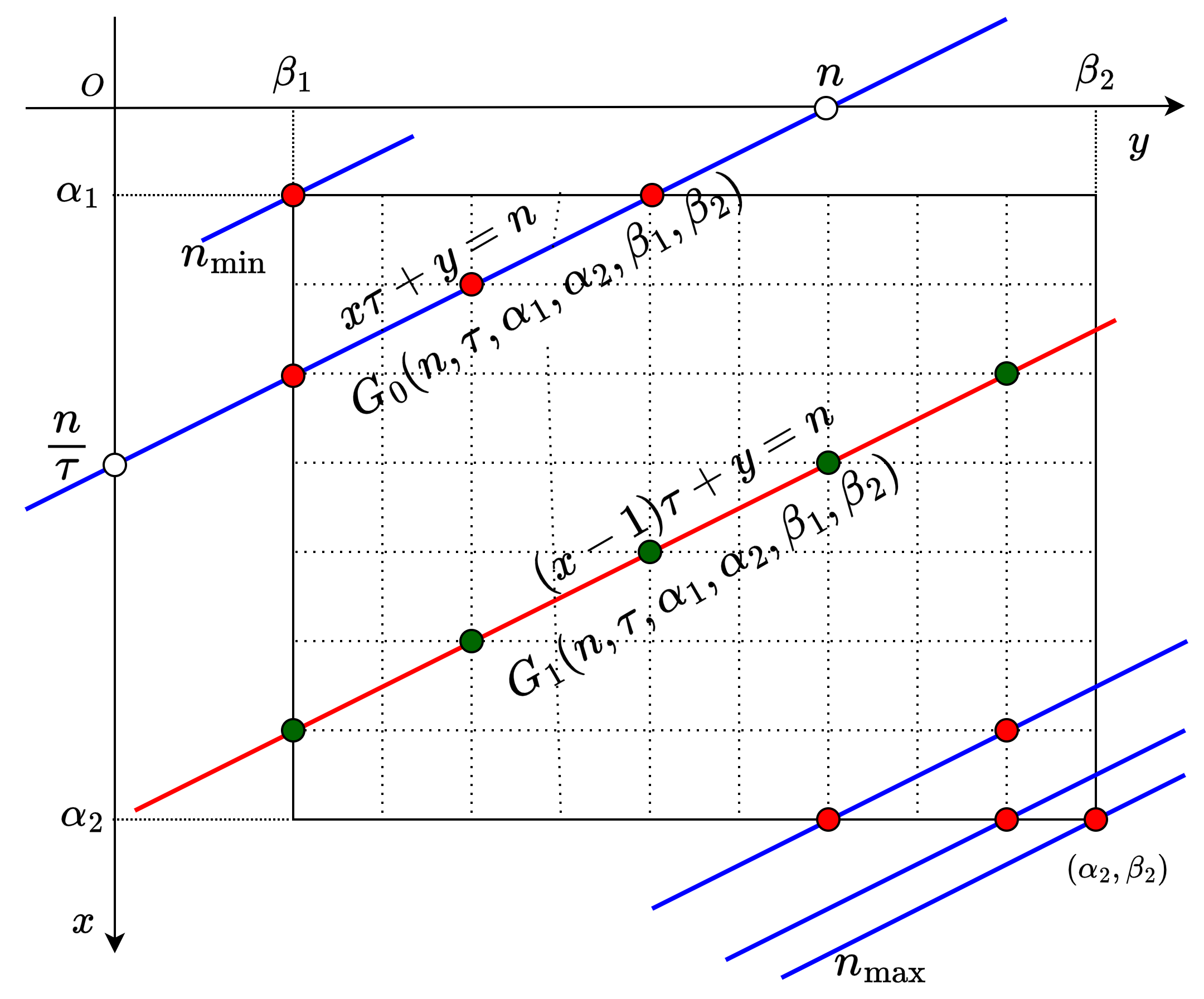} 
\caption{Geometric interpretation of $G_s(n,\tau,\alpha_1, \alpha_2,\beta_1,\beta_2)$} \label{fig-Dipahn-eq-geo}
\end{figure}

Suppose $N, d, \tau\in \mathbb{N}$ are positive integers. Suppose that $\alpha_1, \alpha_2, \beta_1, \beta_2\in \mathbb{Z}^+$ such that $0\le \alpha_1 < \alpha_2$ and $0\le \beta_1<\beta_2$. Let
\begin{equation}
\begin{aligned}
\Omega = \Omega(\alpha_1, \alpha_2, \beta_1, \beta_2)
 =\set{(x,y)\in \mathbb{Z}^2: \alpha_1 \le x \le \alpha_2, \beta_1\le y \le \beta_2}
\end{aligned}
\end{equation}
 For the given $n\in \mathbb{Z}^+$ and $\tau\in \mathbb{N}$,  we have $\GCD(\tau,1) = 1$ and $n> \tau \cdot 1 -\tau - 1 = -1$, thus the set of solutions to the Diophantine equation 
\begin{equation} \label{eq-xy-diop}
\tau x + y = n+s\tau, \quad \forall n\in \mathbb{Z}^+
\end{equation}
must be non-empty according to the \Thm  \ref{thm-appendix-1}  and \Thm  \ref{thm-appendix-2} in \textbf{Appendix} \ref{appendix-1}. For $\forall n \in \mathbb{Z}_+$, let
	\begin{equation}
	\begin{aligned}
	G_s(n,\tau,\alpha_1, \alpha_2,\beta_1,\beta_2) &= \set{(x, y)\in \Omega: x\tau + y = n + s\tau} 
	\end{aligned}
	\end{equation}
	for $s\in \set{0, 1}$. 
The cardinality of $G_s(n,\cdots)$ is denoted by $\abs{G_s(n, \cdots)}$. The geometric interpretation of the set $G_s(n,\cdots)$ is all of the 2-dim points which satisfy \eqref{eq-xy-diop} with interger coordinates on the line $x\tau +y = n+s\tau$ and in the rectangle domain $\Omega$. \Fig \ref{fig-Dipahn-eq-geo} illustrates the scenario intuitively.

The equation $x\tau +y = n+s\tau$ implies that $y = n+s\tau - x\tau$. Thus $\beta_1\le y\le \beta_2$ implies that 
\begin{equation}
\frac{n+s\tau-\beta_2}{\tau} \le x \le \frac{n+s\tau-\beta_1}{\tau}
\end{equation}
It is obvious that
\begin{equation} \label{eq-Dioph-sol}
\max\left(\alpha_1, \frac{n+s\tau-\beta_2}{\tau}\right) \le x \le \min\left(\alpha_2, \frac{n+s\tau-\beta_1}{\tau}\right)
\end{equation}
For the purpose of finding integer solutions, the $\dfrac{n+s\tau-\beta_2}{\tau}$ should be replaced by $\mceil{\dfrac{n+s\tau-\beta_2}{\tau}}$ and the $\dfrac{n+s\tau-\beta_1}{\tau}$ should be replaced by $\mfloor{\dfrac{n+s\tau-\beta_1}{\tau}}$, where  $\mceil{x}$ and $\mfloor{x}$ denote the ceiling and floor of $x\in \mathbb{R}$ respectively. 
 Let
\begin{equation} \label{eq-Dioph-x-range}
\left\{
\begin{aligned}
\scrd{x}{min}^s(n,\tau, \alpha_1,\beta_2) &= \max\left(\alpha_1, \mceil{\frac{n+s\tau-\beta_2}{\tau}}\right)\\
\scrd{x}{max}^s(n,\tau,\alpha_2,\beta_1) &= \min\left(\alpha_2, \mfloor{\frac{n+s\tau-\beta_1}{\tau}}\right)
\end{aligned}
\right.
\end{equation}
for $s\in \set{0,1}$, then it is easy to prove the following theorem according to  \eqref{eq-Dioph-sol} and \eqref{eq-Dioph-x-range}. 

\begin{thm} \label{thm-sol-Diophantine}
For the constrained Diophantine equation 
\begin{equation} \label{eq-Dioph-c}
\tau x + y = n + s\tau, \quad (x,y)\in \Omega, s\in \set{0,1}
\end{equation}
the set of its solutions can be written by
\begin{equation}
\begin{aligned}
G_s(n,\tau, \alpha_1, \alpha_2, \beta_1, \beta_2)
 =\set{(x, n-x\tau): \scrd{x}{min}^s(n,\tau, \alpha_1,\beta_2) \le  x  \le \scrd{x}{max}^s(n,\tau,\alpha_2,\beta_1) }
\end{aligned}
\end{equation}
and the number of solutions is
\begin{equation}
\begin{aligned}
\abs{G_s(n,\tau, \alpha_1, \alpha_2, \beta_1, \beta_2)} 
= \scrd{x}{max}^s(n,\tau,\alpha_2,\beta_1) - \scrd{x}{min}^s(n,\tau, \alpha_1,\beta_2)+1.\\
\end{aligned}
\end{equation}
\end{thm}

\subsubsection{Special Cases}

There are some typical special cases about the structure of $G_s(n,\tau, \alpha_1, \alpha_2, \beta_1, \beta_2)$ for $s\in \set{0, 1}$, $(\alpha_1, \alpha_2) \in \set{(0, d-1), (1, d)}$ and $(\beta_1, \beta_2)\in \set{(0, m-1), (1, m)}$  because of the way of representation of time sequences and coding with concrete computer programming languages. 

\subsubsection*{A. Case of $(\alpha_1, \alpha_2, \beta_1, \beta_2) = (0, d-1, 0, m-1)$}

Particularly, for $d, m\in \mathbb{N}$ and  $(\alpha_1, \alpha_2, \beta_1, \beta_2) = (0, d-1, 0, m-1)$, we have
\begin{equation} \label{eq-set-G0}
\left\{
\begin{aligned}
G_0(n,\tau, 0, d-1, 0, m-1) 
&= \left\{(x, n-x\tau): \scrud{x}{0}{min}(n,\tau, 0, m-1) \le x \le \scrud{x}{0}{max}(n,\tau, d-1,0)  \right\} \\
\abs{G_0(n,\tau,0, d-1, 0, m-1)} 
&= \scrud{x}{0}{max}(n,\tau,d-1,0)  - \scrud{x}{0}{min}(n,\tau,0,m-1)+1
\end{aligned}
\right.
\end{equation}
As an illustration, for the parameter configuration $(N, \tau, d,  m) = (27, 3, 7,  9)$ and $(\alpha_2,\alpha_2, \beta_1, \beta_2) = (0, 6, 0, 8)$, we can obtain 
\begin{equation}
\left\{
\begin{aligned}
&\scrud{x}{0}{min}(n,3,0,8) = \max\left(0, \mceil{\frac{n-8}{3}}\right) \\
&\scrud{x}{0}{max}(n,3,6,0) = \min\left(6, \mfloor{\frac{n}{3}}\right) \\
\end{aligned}
\right.
\end{equation}
and
\begin{equation}
\begin{aligned}
G_0(n,3,0,6,0,8) 
= \set{(x,n-3x): \scrud{x}{0}{min}(n,3,0,8) \le x \le \scrud{x}{0}{max}(n,3,6,0)}
\end{aligned}
\end{equation}
The structure of $G_0(n,\tau, 0, d-1, 0, m-1)= G_0(n,3,0, 6, 0, 8)$ is listed in \Tab \ref{tab-Dioph-eg-sol}. 

\begin{table}[H]
\centering
\caption{Structure of $G_0=\set{(x,n-3x):  \scrud{x}{0}{min} \le x \le \scrud{x}{0}{max}}$ for $N=27$, $\tau=3$, $d=7$, $m=9$, $\scrud{x}{0}{min}= \max\left(0, \mceil{\frac{n-8}{3}}\right)$ and $\scrud{x}{0}{max}=\min\left(6, \mfloor{\frac{n}{3}}\right)$} \label{tab-Dioph-eg-sol}
\begin{tabular}{ccclc}
\hline
$n$ & $\scrud{x}{0}{min}$ & $\scrud{x}{0}{max}$  & $G_0$  &  $\abs{G_0}$  \\
\hline
 $0$ & $0$ & $0$ & $\set{(0,0)}$  & $1$ \\
$1$ & $0$ &  $0$ & $\set{(0,1)}$  & $1$ \\
$2$ & $0$ & $0$ & $\set{(0,2)}$  & $1$\\
$3$ & $0$ & $1$ & $\set{(0,3), (1,0)}$ & $2$\\
$4$ & $0$ & $1$ & $\set{(0,4), (1,1)}$  & $2$ \\
$5$ & $0$ & $1$ & $\set{(0,5), (1,2)}$  & $2$ \\
$6$ & $0$ & $2$ & $\set{(0,5), (1,2), (2, 0)}$  & $3$ \\
$7$ & $0$ & $2$ & $\set{(0,7), (1,4), (2, 1)}$  & $3$ \\
$8$ & $0$ & $2$ & $\set{(0,8), (1,5), (2, 2)}$  & $3$ \\
$9$ & $1$ & $3$ & $\set{(1,6), (2,3), (3, 0)}$  & $3$ \\
$10$& $1$ & $3$ & $\set{(1,7), (2,4), (3, 1)}$  & $3$ \\
$11$& $1$ & $3$ & $\set{(1,8), (2,5), (3, 2)}$  & $3$ \\
$12$& $2$ & $4$ & $\set{(2,6), (3,3), (4, 0)}$  & $3$ \\ 
$13$& $2$ & $4$ & $\set{(2,7), (3,4). (4, 1)}$  & $3$ \\
$14$& $2$ & $4$ & $\set{(2,8), (3,5), (4, 2)}$  & $3$ \\
$15$ & $3$ & $5$ & $\set{(3,6), (4, 3), (5, 0)}$ & $3$\\
 $\vdots$ &  $\vdots$ &  $\vdots$ & \quad $\vdots$  & $\vdots$ \\
$26$ & $6$& $6 $  & $\set{(6,8)}$ & $1$\\
\hline
\end{tabular}
\end{table}

\subsubsection*{B. Case of $(\alpha_2,\alpha_2, \beta_1, \beta_2) = (1, d, 1, m)$}
 
Particularly, for $d, m\in \mathbb{N}$ and  $(\alpha_1, \alpha_2, \beta_1, \beta_2) = (1, d, 1, m)$, we can find that 
\begin{equation} \label{eq-set-G1}
\left\{
\begin{aligned}
G_1(n,\tau, 1, d, 1, m) 
&= \left\{(x, n+\tau-x\tau): \scrud{x}{1}{min}(n,\tau, 1, m) \le x   \le \scrud{x}{0}{max}(n,\tau, d,1)  \right\} \\
\abs{G_1(n,\tau,1, d, 1, m)} 
&= \scrud{x}{1}{max}(n,\tau,d,1)  - \scrud{x}{1}{min}(n,\tau,1,m)+1
\end{aligned}
\right.
\end{equation}
 
Similarly, for the parameter configuration $(N, \tau, d,  m) = (27, 3, 7,  9)$ and $(\alpha_1,\alpha_2, \beta_1, \beta_2) = (1, 7, 1, 9)$, we immediately have 
\begin{equation}
\left\{
\begin{aligned}
&\scrud{x}{1}{min}(n,3,1,9) = \max\left(1, \mceil{\frac{n-6}{3}}\right) \\
&\scrud{x}{1}{max}(n,3,7,1) = \min\left(7, \mfloor{\frac{n+2}{3}}\right) \\
&G_1(n,3,1,7,1,9) 
= \set{(x,n+3-3x): \scrud{x}{1}{min} \le x \le \scrud{x}{1}{max}}
\end{aligned}
\right.
\end{equation}
and the structure of $G_1(n,\tau, 1, d, 1, m)= G_1(n,3, 1, 7, 1, 9)$ is listed in \Tab \ref{tab-Dioph-eg-sol-2}.
\begin{table}[H]
\centering
\caption{Structure of $G_1=\tiny{\set{(x,n+3-3x):  \tiny{\scrud{x}{1}{min} \le x \le \scrud{x}{1}{max}}}}$ for $N=27$,$\tau=3$,$d=7$,$m=9$, $\scrud{x}{1}{min}= \max\left(1, \mceil{\frac{n-6}{3}}\right)$ and $\scrud{x}{1}{max}=\min\left(7, \mfloor{\frac{n+2}{3}}\right)$.} \label{tab-Dioph-eg-sol-2}
\begin{tabular}{ccclc}
\hline
$n$ & $\scrud{x}{1}{min}$ & $\scrud{x}{1}{max}$  & $G_1$  &  $\abs{G_1}$  \\
\hline
 $1$ & $1$ & $1$ & $\set{(1,1)}$  & $1$ \\
$2$ & $1$ &  $1$ & $\set{(1,2)}$  & $1$ \\
$3$ & $1$ & $1$ & $\set{(1,3)}$ & $1$\\
$4$ & $1$ & $2$ & $\set{(1,4), (2,1)}$  & $2$ \\
$5$ & $1$ & $2$ & $\set{(1,5), (2,2)}$  & $2$ \\
$6$ & $1$ & $2$ & $\set{(1,6), (2,3)}$  & $2$ \\
$7$ & $1$ & $3$ & $\set{(1,7), (2,4), (3, 1)}$  & $3$ \\
$8$ & $1$ & $3$ & $\set{(1,8), (2,5), (3, 2)}$  & $3$ \\
$9$ & $1$ & $3$ & $\set{(1,9), (2,6), (3, 3)}$  & $3$ \\
$10$& $2$ & $4$ & $\set{(2,7), (3,4), (4, 1)}$  & $3$ \\
$11$& $2$ & $4$ & $\set{(2,8), (3,5), (4, 2)}$  & $3$ \\
$12$& $2$ & $4$ & $\set{(2,9), (3,6), (4, 3)}$  & $3$ \\ 
$13$& $3$ & $5$ & $\set{(3,7), (4,4). (5, 1)}$  & $3$ \\
$14$& $3$ & $5$ & $\set{(3,8), (4,5), (5, 2)}$  & $3$ \\
$15$ & $3$ & $5$ & $\set{(3,9), (4, 6), (5, 3)}$ & $3$\\
$16$ & $4$ & $6$ & $\set{(4,7), (5, 4), (6, 1)}$ & $3$\\
 $\vdots$ &  &  & \quad $\vdots$  & $\vdots$ \\
$27$ & $7$& $7 $  & $\set{(7,9)}$ & $1$\\
\hline
\end{tabular}
\end{table}

\subsubsection*{C. Case of $(\alpha_1,\alpha_2,\beta_1,\beta_2)=(1,d,1, m)$ and  $\tau=1$}
\begin{figure}[h]
\centering
\includegraphics[width=0.7\textwidth]{ 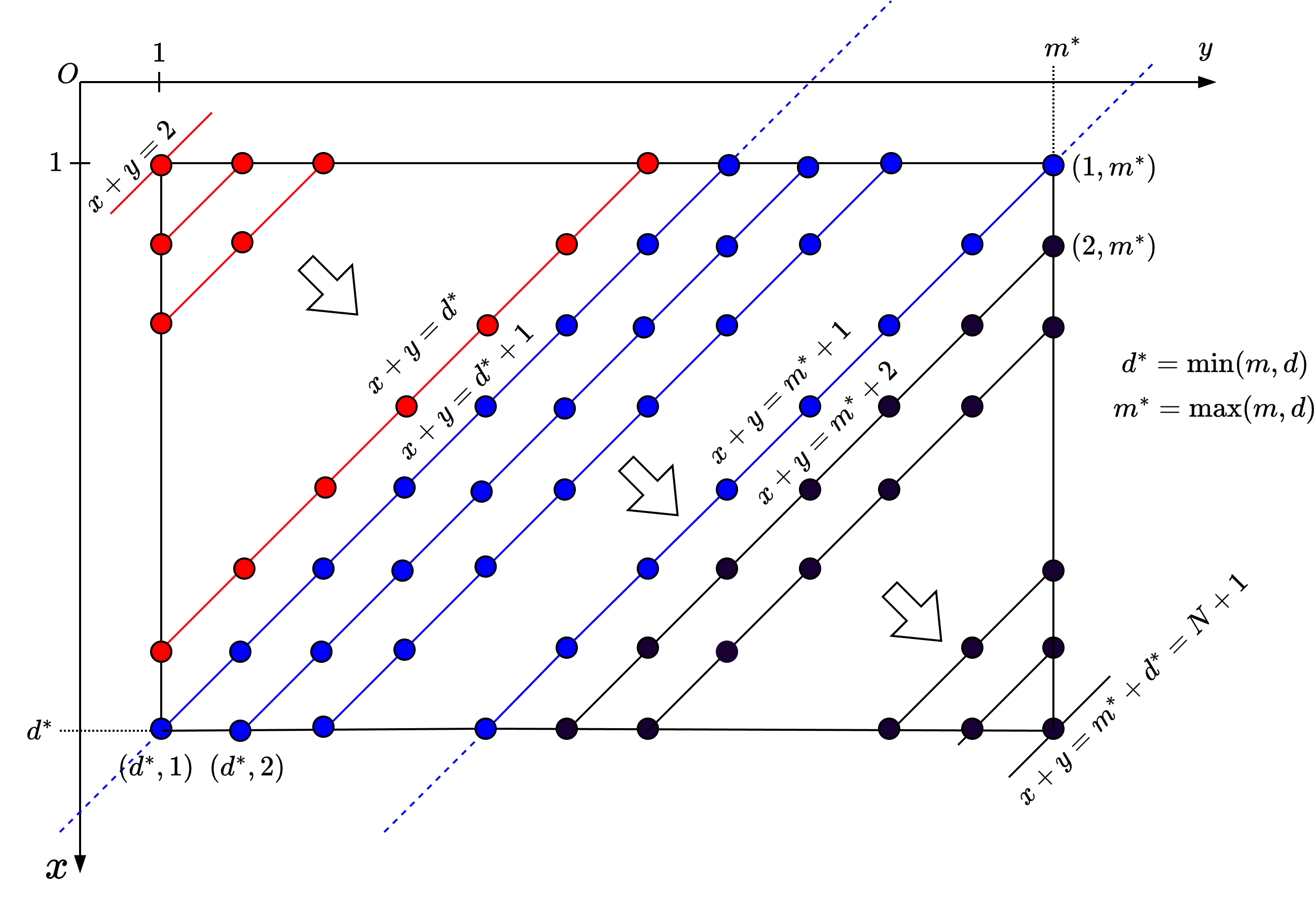} 
\caption{Geometric interpretation of $G_1(n, 1, 1, d, 1, m)$} \label{fig-G1}
\end{figure} 

A special case for the $G_1(n,\tau,\alpha_1,\alpha_2,\beta_1,\beta_2)$ such that $(\tau,\alpha_1,\alpha_2,\beta_1,\beta_2) = (1,1,d,1,m)$ is of interest in the references. \Fig  \ref{fig-G1} illustrates such a scenario.  Let
\begin{equation}
d^* = \min(d, m), \quad m^* = \max(d, m)
\end{equation}
then $d + m = \min(d,m) + \max(d, m)=d^*+m^*$. Thus $N = m + (d-1)\tau = m + d - 1 = d^* + m^* -1$ or equivalently $d^*+m^*=N+1$. For the set
\begin{equation}
\begin{aligned}
G_1^*(n) 
= G_1(n,1,1,d,1,m) 
= \set{(x,n+1-x): \scrud{x}{1}{min}(n,1,1,m)\le x\le \scrud{x}{1}{max}(n,1,d,1)},
\end{aligned}
\end{equation}
we can obtain some interesting observations by \Fig  \ref{fig-G1}. Actually, we have:
\begin{itemize}
\item[\ding{172}] For $1\le n< d^*$, there are $n$ solutions to the Diophantine equation $\tau x + y = n + \tau $ for $(x,y)\in \Omega(1,d,1,m)$ and $s=1$, thus
\begin{equation} \label{eq-G1-1}
\left\{
\begin{aligned}
&G_1^*(n) 
= \left\{(x,n+1-x): 1\le x\le n\right\} \\
&\abs{G_1^*(n)} = n
\end{aligned}
\right. 
\end{equation}
Geometrically, the $G_1^*(n)$ corresponds to the red points on the red hypotenuse of the equilateral right angled triangle  shown in the \Fig  \ref{fig-G1}.
\item[\ding{173}] For $d^* \le n \le m^*$, there are always $d^*$ solutions to  the   Diophantine equation
$\tau x + y = n + s\tau$ for $(x,y)\in \Omega(1,d,1,m)$, thus
\begin{equation} \label{eq-G1-2}
\left\{
\begin{aligned}
&G_1^*(n) = \left\{(x, n+1-x): 1+ \max(0,n-m)\le x  \le \min(d,n)\right\},  \\
&\abs{G_1^*(n)} = d^*
\end{aligned}
\right. 
\end{equation} 
Geometrically, the $G_1^*(n)$ corresponds to the blue points on the blue edge of the parallelogram  shown in the \Fig  \ref{fig-G1}.
\item[\ding{174}] For $m^* < n \le N$, there are $N-n+1$ solutions to  the  Diophantine equation
$\tau x + y = n + s\tau$ for $(x,y)\in \Omega(1,d,1,m)$, thus
\begin{equation} \label{eq-G1-3}
\left\{
\begin{aligned}
G_1^*(n) 
&= \set{(n+1-y, y): n-m^* +1\le y  \le N-m^* +1}  \\
&= \set{(x, n+1-x): m^* +n - N \le x \le m^*}\\
\abs{G_1^*(n)} &= N-n+1
\end{aligned}
\right. 
\end{equation} 
for $m^* < n\le N$.
Geometrically, the $G_1^*(n)$ corresponds to the black points on the black edge of the parallelogram  shown in the \Fig  \ref{fig-G1}.
\end{itemize}

\section{Embedding of Time Sequence} \label{sect-embedding-ts}

\subsection{General Principle}
The time-delay embedding theorem proposed by Floris Takens \cite{Robinson2010,Takens1981} based on time sequence delay topology shows that any 1-dim time sequence
can be converted to a trajectory matrix --- multi-dimensional time sequence matrix. Suppose that $\tau\in \mathbb{N}$ is the time delay and $d$ is the guessed embedding dimension such that $d\ge 2\scrd{n}{dof} + 1$. Let
	\begin{equation}
    m = N- (d-1)\tau 
    \end{equation}	
be the numbers of$d$-dim signals, then the noisy $d$-by-$m$ trajectory matrix $\mat{M}\in \ES{R}{d}{m}$  can be expressed by	
\begin{equation}
\begin{aligned}
\Phi_{s,\tau}: \mathscr{X} &\to \ES{R}{d}{m} \\
X&\mapsto \mat{M}=(M^{i+s}_{j+s})_{d\times m}
\end{aligned}
\end{equation}
such that
\begin{equation} \label{eq-X-traj-mat}
	\begin{split}
		\mat{M}
		&= \Phi_{s,\tau}(X) 
		= [\bf{x}_s, \bf{x}_{s+1}, \cdots, \bf{x}_{s+m-1}]
		=  \left(M^{i+s}_{j+s}\right)_{d\times m} =\left(x[i\tau+j+s] \right)_{d\times m} \\
		&= \begin{bmatrix}
			x[s] & x[s+ 1]  & \cdots & x[s+m-1]  \\
			x[\tau+s] & x[\tau+s+1]  & \cdots & x[\tau +s+m-1] \\
			\vdots & \vdots   & \ddots & \vdots \\
			x[(d-1)\tau+s] & x[(d-1)\tau+s+1] & \cdots & x[(d-1)\tau+s+m-1]
		\end{bmatrix}
\end{split}
\end{equation}
for $0\le i \le d-1$ and $0 \le j \le m-1$, where $s\in \set{0, 1}$ for the time sequence $X=\seq{x[s], x[s+1], \cdots, x[N-1+s]}$. 

Please note that different value of $\tau$ will lead to different size of the trajectory matrix $\mat{M}$, thus the computational complexity of decomposing $\mat{M}$ will be much different. \Tab \ref{tab-size-mat-M} illustrates the size of $\mat{M}\in \ES{R}{d}{m}$ for $(N,d)=(2000,100)$.  If the length  $N$ pf the time sequence $X$ is small in applications, we can set the time delay with a default value $\tau = 1$. 

\begin{table}[h]
\centering
\caption{Size of the trajectory matrix $\mat{M}\in \ES{R}{d}{m}$}
\label{tab-size-mat-M}
\begin{tabular}{cccclc}
\hline
$N$ & $d$ & $\tau$ & $m$ & $d\times m$ & Remark \\
\hline
2000 & 100 & 1 & 1901 & $100\times 1901$ & $ m\ge d$ \\
2000 & 100 & 2 & 1802 & $100\times 1802$ & $ m\ge d$\\
2000 & 100 & 3 & 1703 & $100\times 1703$ & $ m\ge d$\\
2000 & 100 & 4 & 1604 & $100\times 1604$ & $ m\ge d$\\
2000 & 100 & 5 & 1505 & $100\times 1505$ & $ m\ge d$\\
2000 & 100 & 6 & 1406 & $100\times 1406$ & $ m\ge d$\\
2000 & 100 & 7 & 1307 & $100\times 1307$ & $ m\ge d$\\
2000 & 100 & 8 & 1208 & $100\times 1208$ & $ m\ge d$\\
2000 & 100 & 9 & 1109 & $100\times 1109$ & $ m\ge d$\\
2000 & 100 & 10 & 1010 & $100\times 1010$ & $ m\ge d$\\
2000 & 100 & 11 & 911 & $100\times 911$ & $ m\ge d$\\
2000 & 100 & 12 & 812 & $100\times 812$ & $ m\ge d$\\
2000 & 100 & 13 & 713 & $100\times 713$ & $ m\ge d$\\
2000 & 100 & 14 & 614 & $100\times 614$ & $ m\ge d$\\
2000 & 100 & 15 & 515 & $100\times 515$ & $ m\ge d$\\
2000 & 100 & 16 & 416 & $100\times 416$ & $ m\ge d$\\
2000 & 100 & 17 & 317 & $100\times 317$ & $ m\ge d$\\
2000 & 100 & 18 & 218 & $100\times 218$ & $ m\ge d$\\
2000 & 100 & \fbox{19} & \fbox{119} & $100\times 119$ & $ m\ge d$\\
2000 & 100 & 20 & 20  & $100\times 20$ & $ m< d$\\
\hline
\end{tabular}
\end{table}

Particularly, for $s = 0$, we have the trajectory matrix of type-0 
\begin{equation} \label{eq-X-traj-mat-0}
	\begin{split}
		\mat{M} 
		&=\Phi_{0,\tau}(X, d, m) 
		= [\bf{x}_0, \bf{x}_1, \cdots, \bf{x}_{m-1}]
		=  \left(M^i_j\right)_{d\times m} =\left(x[i\tau+j] \right)_{d\times m}\\
		&= \begin{bmatrix}
			x[  0] & x[1]  & \cdots & x[m-1] \\
			x[\tau] & x[\tau+1]  & \cdots & x[\tau +m-1] \\
			\vdots & \vdots   & \ddots & \vdots \\
			x[(d-1)\tau] & x[(d-1)\tau+1] & \cdots & x[(d-1)\tau+m-1]
		\end{bmatrix}
\end{split}
\end{equation}
for the row index $0\le i \le d-1$ and column index $0\le j\le m-1$ 
where $\bf{x}_j = \trsp{\begin{bmatrix}x[j], x[\tau+j], \cdots, x[(d-1)\tau+j]\end{bmatrix}}$  is a $d$-dim vector. It is easy to find that 
 \begin{itemize}
 \item $ M^0_0 = x[0]$ and $M^{d-1}_{m-1} = x[(d-1)\tau + m-1] = x[N-1]$;
 \item each of the $x[n]$ for $0\le n\le N-1$ has been embedded in the trajectory matrix;
 \item for the given $n$, the $x[n]$ appears at the position of $(x,y)$ such that $\tau x + y = n$ for $0\le x\le d-1$ and  $0\le y\le m-1$.
\end{itemize}

Similarly, for $s = 1$, we have the trajectory matrix of type-1 as follows
\begin{equation} \label{eq-X-traj-mat-1}
	\begin{split}
		\mat{M} 
		&=\Phi_{1,\tau}(X,d, m) 
		= [\bf{x}_1, \bf{x}_2, \cdots, \bf{x}_{m}]
		=  \left(M^i_j\right)_{d\times m} =\left(x[(i-1)\tau+j] \right)_{d\times m}\\
		&=\begin{bmatrix}
			x[1] & x[2]  & \cdots & x[m] \\
			x[\tau+1] & x[\tau+2]  & \cdots & x[\tau +m] \\
			\vdots & \vdots   & \ddots & \vdots \\
			x[(d-1)\tau+1] & x[(d-1)\tau+2] & \cdots & x[(d-1)\tau+m]
		\end{bmatrix}
\end{split}
\end{equation}
for $1\le i \le d$ and  $1\le j\le m$ 
where 
$$\bf{x}_{j} = \trsp{\begin{bmatrix}x[j], x[\tau+j], \cdots, x[(d-1)\tau+j]\end{bmatrix}}$$
  is a $d$-dim signal vector. It is easy to find that 
 \begin{itemize}
 \item $ M^1_1 = x[1]$ and $X^d_m = x[(d-1)\tau + m] = x[N]$;
 \item each of the $x[n]$ for $1\le n\le N$ has been embedded into the trajectory matrix;
 \item for given $n$, the $x[n]$ appears at the position of $(i,j)$ such that $\tau (i-1) + j = n$ for $1\le i\le d$ and
 $1\le j\le m$.
\end{itemize}

\subsection{Examples and Interpretations}

\Fig  \ref{fig-X-trajmat-python} demonstrates the embedding of the sequence $X=\seq{x[0], x[1], \cdots, x[26]}$ of type-0 into the trajectory matrix $\mat{M}=(M^i_j)\in \ES{R}{7}{9}$ with the parameter configuration $(N, d, \tau, m) = (27, 7, 3, 9)$.
 \begin{figure*}[htbp]
 \centering
 $$
\begin{bmatrix}
 x[0] & x[1] & x[2] & x[3] & x[4]& x[5] & x[6] & x[7]& x[8] \\
 x[3] & x[4] & x[5] & x[6] & x[7] & x[8] & x[9] & x[10] & x[11] \\
 x[6] & x[7] & x[8] & x[9] & x[10] & x[11] &x[12] & x[13] & x[14]  \\
 x[9] & x[10] & x[11] &x[12] & x[13] & [14] &x[15] & x[16] & x[17]   \\
 x[12] & x[13] & [14] &x[15] & x[16] & x[17] & x[18] & x[19] & x[20]  \\
 x[15] & x[16] & x[17] & x[18] & x[19] & x[20] & x[21] & x[22] & x[23] \\
 x[18] & x[19] & x[20] & x[21] & x[22] & x[23] & x[24] & x[25] & x[26] \\
 \end{bmatrix}
 $$
 \caption{Convert the sequence $X=\seq{x[0], x[1], \cdots, x[26]}$ to the trajectory matrix $\mat{M}$ with $(N, d, \tau, m) = (27, 7, 3, 9)$ }
 \label{fig-X-trajmat-python}
 \end{figure*}
 
 We now give some necessary interpretations about correspondence of the \Tab \ref{tab-Dioph-eg-sol} and \Fig  \ref{fig-X-trajmat-python} as follows:
\begin{itemize}
\item For $n=0$ in \Tab \ref{tab-Dioph-eg-sol}, we have $G_0(0, \cdots)=\set{(0,0)}$, which means the element $x[0]$ of time sequence $X$ appears as the entry $M^0_0$ (located in the $0$-th row and $0$-th column) in \Fig  \ref{fig-X-trajmat-python}. In other words,  $x[0]$ appears only one time in the matrix $\mat{M}=(M^i_j)_{7\times 9}$ with position indicated by the set $G_0(0, \cdots)=\set{(0,0)}$. 
\item  For $n=1$ in \Tab \ref{tab-Dioph-eg-sol}, we have $G_0(0, \cdots)=\set{(0,1)}$, which means the element $x[1]$ of time sequence $X$ appears as the entry $M^0_1$ (located in the $0$-th row and $1$-th column) in \Fig  \ref{fig-X-trajmat-python}. In other words,  $x[1]$ appears only one time in the matrix $\mat{M}=(M^i_j)_{7\times 9}$ with position indicated by the set $G_0(1, \cdots)=\set{(0,1)}$. 
\item \quad $\vdots$
\item For $n=8$ in \Tab \ref{tab-Dioph-eg-sol}, we can find that $G_0(8, \cdots)=\set{(0,8), (1, 5), (2, 2)}$, which means the element $x[8]$ appears as the entries $M^0_8, M^1_5$ and $M^2_2$  in \Fig  \ref{fig-X-trajmat-python}. In other words,  $x[8]$ appears three times in the matrix $\mat{M}=(M^i_j)_{7\times 9}$ with positions indicated by the set $G_0(n, \cdots)$.
\item \quad $\vdots$
\end{itemize}

\Fig  \ref{fig-X-trajmat-matlab} demonstrates the embedding of the sequence $X=\seq{x[1], x[1], \cdots, x[27]}$ of type-1 into the trajectory matrix $\mat{M}=(M^i_j)\in \ES{R}{7}{9}$ with the parameter configuration $(N, d, \tau, m) = (27, 7, 3, 9)$. The verification of the correspondence of \Tab \ref{tab-Dioph-eg-sol-2} and \Fig \ref{fig-X-trajmat-matlab} is trivial and we omitted it here.
 \begin{figure*}[htbp]
 \centering
$$
 \begin{bmatrix}
 x[1] & x[2] & x[3] & x[4]& x[5] & x[6] & x[7]& x[8] & x[9] \\
 x[4] & x[5] & x[6] & x[7] & x[8] & x[9] & x[10] & x[11] & x[12]\\
 x[7] & x[8] & x[9] & x[10] & x[11] &x[12] & x[13] & x[14] & x[15] \\
 x[10] & x[11] &x[12] & x[13] & [14] &x[15] & x[16] & x[17] & x[18]  \\
 x[13] & [14] &x[15] & x[16] & x[17] & x[18] & x[19] & x[20]  & x[21]\\
 x[16] & x[17] & x[18] & x[19] & x[20] & x[21] & x[22] & x[23] & x[24]\\
 x[19] & x[20] & x[21] & x[22] & x[23] & x[24] & x[25] & x[26] & x[27] \\
 \end{bmatrix}
 $$
 \caption{Convert the sequence $X=\seq{x[1], x[2], \cdots, x[27]}$ to the trajectory matrix $\mat{M}$ with $(N, d, \tau, m) = (27, 7, 3, 9)$ }
 \label{fig-X-trajmat-matlab}
 \end{figure*}
 
\section{Pulling Back of Immersion Space} \label{sect-pullback}

\subsection{Method and Steps}
 
The fact that the element $x[n]$ of the TS $X$ appears on the line with positions denoted by $G_s(n, \tau, \alpha_1, \alpha_2, \beta_1, \beta_2)$ demonstrated in \Fig \ref{fig-Dipahn-eq-geo} can be used to rebuild the element $x[n]$ from the trajectory matrix $\mat{M}$. For this purpose, what we should do is just averaging the entries of the matrix $\mat{M}$ labeled by the set $G_s(n, \tau, \alpha_1, \alpha_2, \beta_1, \beta_2)$. There are four simple steps:
 \begin{itemize}
 \item firstly, specify the parameters $s\in \set{0,1}$, $N\in \mathbb{N}$, $\tau\in \mathbb{N}$, $d\in \mathbb{N}$, $m\in \mathbb{N}$  and the concrete form of the trajectory matrix $\mat{M}\in \ES{R}{d}{m}$ by \eqref{eq-X-traj-mat-0} or \eqref{eq-X-traj-mat-1};
 \item secondly, computing the range parameter $\scrd{x}{min}^s(n,\tau,\alpha_1,\beta_2)$ and $\scrd{x}{max}^s(n,\tau,\alpha_2,\beta_1)$ for the set  $G_s(n,\tau,\alpha_1,\alpha_2,\beta_1,\beta_2)$;
 \item thirdly, computing the number $\abs{G_s(n, \tau, \alpha_1, \alpha_2, \beta_1, \beta_2)}=\scrd{x}{max}^s(n,\tau,\alpha_2,\beta_1) - \scrd{x}{min}^s(n,\tau,\alpha_1,\beta_2) + 1$;
 \item finally, averaging the entries of $\mat{M}$ according to the set $G_s(n, \tau, \alpha_1, \alpha_2, \beta_1, \beta_2)$.
 \end{itemize}

\subsection{Pulling Back Theorem }

Suppose the trajectory matrix $\mat{M}\in \ES{R}{d}{m}$ is decomposed into a group of matrices $\mat{Z}_1, \mat{Z}_2, \cdots, \mat{Z}_r$ such that
\begin{equation}
\mathcal{D}(\mat{M}) = \sum^r_{k=1}\mat{Z}_k
\end{equation}
where $\mat{Z}_k = (Z^i_j(k))_{d\times m}$  is the $k$-th CTM and $\mathcal{D}$ is the matrix decomposing operation. 

According to the method and steps discussed above, we can deduce the following theorem for the inverse of embedding mapping, which map the $k$-th CTM $\mat{Z}(k)\in \ES{R}{d}{m}$ in the immersion space to the time sequence $X^{(k)} = \set{x^{(k)}[n]}^{N-1+s}_{n=s}$ in the sequence space $\mathscr{X}$:  

\begin{thm}[Noise Free] \label{thm-pullback-noisyfree}
For the type $s\in\set{0,1}$, embedding dimension $d\in \mathbb{N}$ and time sequence $X=\set{x[n]: s\le n \le N-1+s}$ of length $N$, let
\begin{equation} \label{eq-qmin-qmax}
\left\{
\begin{aligned}
\scrd{q}{min} &= \max\left(s, \mceil{\frac{n+s\tau -m + (1-s)}{\tau}} \right) \\
\scrd{q}{max} &= \min \left(d+s-1, \mfloor{\frac{n+(\tau-1)s}{\tau}}\right)
\end{aligned}
\right.
\end{equation}
and
\begin{equation} \label{eq-Q-nstau}
Q_k(n,\tau,s) =\set{Z^q_{n+s\tau -qs}(k): \scrd{q}{min}\le s\le \scrd{q}{max}}
\end{equation}
be the data set which allows duplicate elements for the entries of $\mat{Z}_k\in \ES{R}{d}{m}$
then we can pull back the $k$-th CTM $\mat{Z}_k=(Z^i_j(k))_{d\times m}\in \ES{R}{d}{m}$ in the immersion space to the $k$-th CTS $X^{(k)}=\set{x^{(k)}[n]}^{N-1+s}_{n=s}$ in the sequence space $\mathscr{X}$ by  
\begin{equation} \label{eq-xn-rebuild}
\begin{aligned}
x^{(k)}[n] &= \ProcName{AriAveSolver}(Q_k(n,\tau,s), \scrd{q}{max} - \scrd{q}{min} + 1)\\
&= \frac{1}{\scrd{q}{max} - \scrd{q}{min} + 1} \sum^{\scrd{q}{max}}_{q=\scrd{q}{min}}Z^q_{n+s\tau - q\tau}(k).
\end{aligned}
\end{equation}
where \ProcName{AriAveSolver} the algorithm for solving the arithmetic average.
\end{thm}   

\noindent \textbf{Proof}: 

In order to find the inverse $\inv{\Phi}_{s,\tau}: \ES{R}{d}{m}\to \mathscr{X}$ for the embedding $\Phi_{s,\tau}: \mathscr{X}=\set{x[n]}^{N-1+s}_{n=s}\to \ES{R}{d}{m}, X\mapsto \mat{M}$ and decomposition $\displaystyle \mat{M}=\sum^r_{k=1}\mat{Z}_k$, what we need is to find the position of $x^{(k)}[n]$ appearing in the CTM $\mat{Z}_k$.  

According to the definition of the trajectory matrix $\mat{M}=\sum^r_{k=1}\mat{Z}_k$ of type-0 in \eqref{eq-X-traj-mat-0} or of type-1 in  \eqref{eq-X-traj-mat-0}, it is equivalent to find the solution $(i, j)$ to the Diophantine equation
$\tau i + j = n$ or $\tau i + j = n +\tau$ for the given $n$. Obviously, the key issue lies in solving the binary Diophantine equation $\tau x + y = n + s\tau$ for $s\in \set{0, 1}$. According to the \Thm \ref{thm-sol-Diophantine}, the pair of row and column indices $(i,j)$ are the elements of $G_s(n,\tau, \alpha_1, \alpha_2, \beta_1, \beta_2)$ for $(s, \alpha_1, \alpha_2, \beta_1, \beta_2) = (0, 0, d-1, 0, m-1)$ or $(s, \alpha_1, \alpha_2, \beta_1, \beta_2) = (1, 1, d, 1, m)$ with the form $(i,j) = (q, n+s\tau -q\tau)$ for $ \scrd{q}{min} \le q\le \scrd{q}{max}$.      

On the other hand, the data set $Q_k(n,\tau,s)$ contains all of the candidates or copies of $x^{(k)}[n]$ appearing in the matrix $\mat{Z}_k$. Consequently, the $x^{(k)}[n]$ can be rebuilt  by averaging all of the elements in the data set $Q_k(n,\tau,s)$, which is computed by \eqref{eq-xn-rebuild}. This completes the proof. $\blacksquare$

Particularly, for $\tau=1$ and $s=1$, we have the following corollary by  \Thm  
\ref{thm-pullback-noisyfree}
\begin{cor}	\label{cor-ZHY-matlab}
	For the given $n\in \set{1, 2, \cdots, N}$ and the $k$-th CTM $\mat{Z}_k = (Z^i_j(k))_{d\times m}$ of type-1 in the immersion space $\ES{R}{d}{m}$, let 
\begin{equation}
\left\{
\begin{aligned}
\scrd{p}{min} &= \max\left(1, n+1-m\right)  \\
\scrd{p}{max} &= \min\left(d,n\right)
\end{aligned}
\right.
\end{equation}	
for $1\le i\le d$ and $1\le j\le m$, we can convert the $k$-th CTM $\mat{Z}_k$ to the $k$-th CTS $X^{(k)}=\set{x^{(k)}[n]}^{N-1+s}_{n=s}\in \mathscr{X}$  by
	\begin{equation}
	x^{(k)}[n] = \frac{1}{\scrd{p}{max}-\scrd{p}{min}+1}\sum^{\scrd{p}{max}}_{p=\scrd{p}{min}}Z^{p}_{n+1-p\tau}(k)
	\end{equation}	
\end{cor}

\section{Discussion} \label{sect-discussion}

\subsection{Specific Scenario vs. General Scenario}

We remark that Corollary  \ref{cor-ZHY-matlab} is equivalent to the DAP in Result \ref{res-Pan} which was taken by Pan et al. \cite{Pan2019sgmd} in SGMD. 
\begin{res}[Diagonal Averaging Principle, DAP]	 \label{res-Pan}
	Suppose that $d$ is the embedding dimension, $m = N-(d-1)\tau$ such that $\tau=1$,  and the $k$-th CTM $\mat{Z}_k = (Z^i_j(k))_{d\times m}$ of type-1 is in the immersion space $\ES{R}{d}{n}$. Let $d^* = \min(m,d)$, $m^* = \max(m,d)$ and 
	\begin{equation}
	\tilde{Z}^i_j(k)  = \begin{cases}
	Z^i_j(k) , & m < d; \\
	Z^j_i(k) , & m \ge d.
	\end{cases}
	\end{equation}
for $1\le i\le d$ and $1\le j\le m$, the $k$-th CTS $x^{(k)}[n]$ rebuilt from the $\mat{Z}$ can be computed by 
	\begin{equation}\label{eq-Pan} 
	x^{(k)}[n]
	= \begin{cases}  
	  \cfrac{1}{n}\sum\limits^{n}_{p=1}\tilde{Z}^p_{n-p+1}(k) , & 1 \le n < d^*;\\
	  \cfrac{1}{d^*}\sum\limits^{d^*}_{p=1} \tilde{Z}^p_{n-p+1}(k), & d^* \le n\le m^* \\
	  \cfrac{1}{N-n+1}\sum\limits_{p=n-m^*+1}^{N-m^*+1}\tilde{Z}^p_{n-p+1}(k), & m^* < n \le N
	\end{cases}		
	\end{equation}
\end{res}
Note that Result \ref{res-Pan} just holds for $\tau=1$ and it can be derived from the pulling back method according to the equations \eqref{eq-G1-1}, \eqref{eq-G1-2} and \eqref{eq-G1-3} about the set $G_1^*(n)$.

We remark that the diagonal averaging strategy is originally proposed by Vautard et al. \cite{Vautard1992SSA} in the singular spectrum analysis (SSA) in 1992 and followed by  Jaime et al. \cite{Jaime2014SSA} and Leles et al. \cite{Leles2018SSA}. In the SSA, the immersion matrix is the specific form of the trajectory matrix such that the time delay is $\tau=1$. In consequence, the formula for  converting the CTM in the immersion space to the CTM in the sequence space with the averaging strategy just holds for the special case $\tau = 1$ and it can not be used as a general method for pulling back a CTM to the corresponding CTS.

\subsection{Pulling Back in Decomposing Time Sequence}

The pulling back theorem can be applied to the SGMD as well as the SSA for signal decomposition. Since the signals can be classified into deterministic signals and random signals, the applications of pulling back to signal decomposition can also be classified into two categories.

\subsubsection{Deterministic Time Sequence}

\begin{figure}[h] 
\centering
\includegraphics[width=0.65\textwidth]{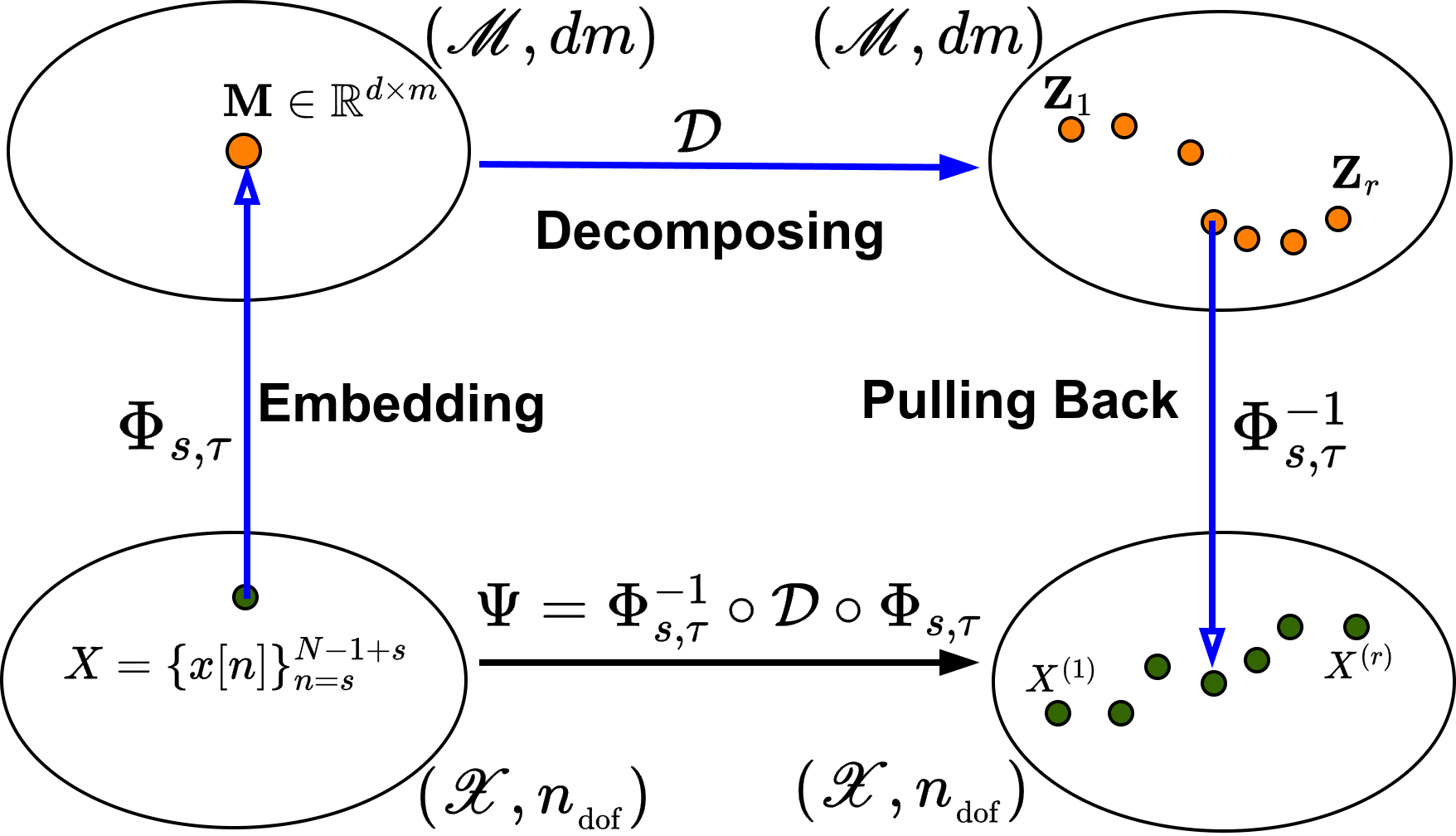} 
\caption{Pulling back theorem for deterministic time sequence}
\label{fig-pullback-deterministic-signal}
\end{figure}

For the deterministic signal $X=\set{x[n]}^{N-1+s}_{n=s}\in \mathscr{X}$, we assume that the decomposition of the corresponding trajectory matrix $\mat{M} = \Phi_{s,\tau}(X)\in \ES{R}{d}{m}$ is given by
\begin{equation}
\sum^r_{k=1}\mat{Z}_k = \mathcal{D}(\mat{M}) = \mathcal{D}\circ \Phi_{s,\tau}(X)
\end{equation}  
in which $(\mat{Z}_1, \mat{Z}_2, \cdots, \mat{Z}_r)$ is a group of CTM. By applying the pulling back theorem to each of the CTM,  we can obtain the corresponding CTS as follows:
\begin{equation}
X^{(k)} = \inv{\Phi}_{s,\tau}(\mat{Z}_k), \quad  1\le k\le r.
\end{equation}
The mode decomposition operation $\Psi: \mathscr{X}\to \mathscr{X}$ can be expressed formally by
\begin{equation}
\begin{aligned}
\Psi: \mathscr{X}&\to \mathscr{X}\\
X &\mapsto \sum^r_{k=1}X^{(k)}=\sum^r_{k=1}\inv{\Phi}_{s,\tau}(\mat{Z}_k)
\end{aligned} 
\end{equation}
In other words, each CTM $Z_k$ will be pulled back to the corresponding 
CTS $X^{(k)}\in \mathscr{X}$. \Fig \ref{fig-pullback-deterministic-signal} illustrates the pulling back of deterministic signal intuitively with commutative diagram via the equivalent mode decomposition operator 
\begin{equation}
\Psi = \inv{\Phi}_{s,\tau}\circ \mathcal{D}\circ \Phi_{s,\tau}, \quad \mathrm{without~denoising}
\end{equation}

\subsubsection{Random Time Sequence}

\begin{figure*}[h]
\centering
\includegraphics[width=\textwidth]{ 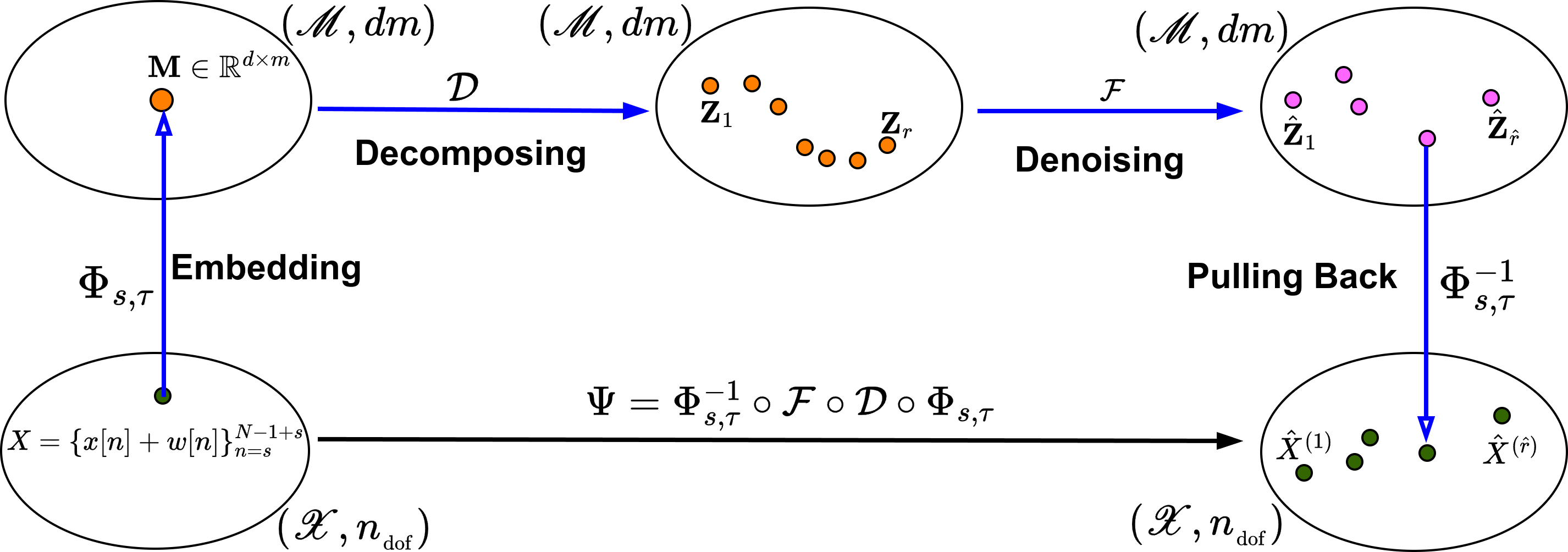} 
\caption{Pulling back theorem for random time sequence}
\label{fig-pullback-random-signal}
\end{figure*}

For practical problems, the time sequence involved is usually random due to the noise in the process of capturing discrete time data with sensor. In order to filter the noise, it is necessary to introduce a denoising module which can be denoted by $\mathcal{F}$. \Fig \ref{fig-pullback-random-signal} illustrates this scenario intuitively.

We remark that the matrix components $\mat{Z}_1, \cdots, \mat{Z}_r$ obtained by the decomposing module are perturbed by noise. After the denoising operation, we have 
\begin{equation}
\sum^{\hat{r}}_{k=1} \hat{\mat{Z}}_k = \mathcal{F}\left(\sum^r_{i=1} \mat{Z}_i\right)
\end{equation}
where $\hat{r}\le r$ since some of the components may be removed and some the components may be modified. If the noise does not exist, then we have $r=\hat{r}$ and $\hat{\mat{Z}}_k = \mat{Z}_k$ or equivalently $\mathcal{F}=\mathbf{1}$ is the identity operator. Generally, we can obtain 
\begin{equation}
\sum^{\hat{r}}_{k=1} \hat{X}^{(k)}=\Psi(X) = (\inv{\Phi}_{s,\tau}\circ \mathcal{F}\circ \mathcal{D}\circ \Phi_{s,\tau})(X)
\end{equation}
where $\hat{X}^{(k)}\in \mathscr{X}$ is the $k$-th estimated component of time sequence
from the perturbed time sequence $X$ with the equivalent mode decomposition operator 
\begin{equation}
\Psi = \inv{\Phi}_{s,\tau}\circ \mathcal{F}\circ \mathcal{D}\circ \Phi_{s,\tau}, \quad \mathrm{with~denoising}
\end{equation}

It should be noted that the noise has important impact on the rebuilding process of $x^{(k)}[n]$ with the container \eqref{eq-Q-nstau}. In the sense of rebuilding the $x^{(k)}[n]$ with $Q_k(n,\tau,s)$, the essence of \eqref{eq-xn-rebuild} is just estimating the sequence $X^{(k)}=\set{x^{(k)}[n]}^{N-1+s}_{n=s}$ with simple arithmetic averaging.

\begin{table}[h]
\centering
\caption{Relation of data set size $\abs{Q_k(n,\tau,s)}$ and time delay $\tau$ for the given $(N, d, s, n)$}
\label{tab-Q-size}
\begin{tabular}{ccccccc}
\hline
$N$ & $d$ & $s$ & $n$ & $\tau$ & $\abs{Q_k(n,\tau,s)}$ & Remark \\
\hline
2000 & 100 & 1 & 213 & 1 & 100 & $\tau\le \scrd{\tau}{max}$ \\
2000 & 100 & 1 & 213 & 2 & 100& $\tau\le \scrd{\tau}{max}$\\
2000 & 100 & 1 & 213 & 3 & 71& $\tau\le \scrd{\tau}{max}$\\
2000 & 100 & 1 & 213 & 4 & 54& $\tau\le \scrd{\tau}{max}$\\
2000 & 100 & 1 & 213 & 5 & 43& $\tau\le \scrd{\tau}{max}$\\
2000 & 100 & 1 & 213 & 6 & 36& $\tau\le \scrd{\tau}{max}$\\
2000 & 100 & 1 & 213 & 7 & 31& $\tau\le \scrd{\tau}{max}$\\
2000 & 100 & 1 & 213 & 8 & 27& $\tau\le \scrd{\tau}{max}$\\
2000 & 100 & 1 & 213 & 9 & 24& $\tau\le \scrd{\tau}{max}$\\
2000 & 100 & 1 & 213 & 10 & 22& $\tau\le \scrd{\tau}{max}$\\
2000 & 100 & 1 & 213 & 11 & 20& $\tau\le \scrd{\tau}{max}$\\
2000 & 100 & 1 & 213 & 12 & 18& $\tau\le \scrd{\tau}{max}$\\
2000 & 100 & 1 & 213 & 13 & 17& $\tau\le \scrd{\tau}{max}$\\
2000 & 100 & 1 & 213 & 14 & 16& $\tau\le \scrd{\tau}{max}$\\
2000 & 100 & 1 & 213 & 15 & 15& $\tau\le \scrd{\tau}{max}$\\
2000 & 100 & 1 & 213 & 16 & 14& $\tau\le \scrd{\tau}{max}$\\
2000 & 100 & 1 & 213 & 17 & 13& $\tau\le \scrd{\tau}{max}$\\
2000 & 100 & 1 & 213 & 18 & 12& $\tau\le \scrd{\tau}{max}$\\
2000 & 100 & 1 & 213 & \fbox{19} & \fbox{7}& $\tau\le \scrd{\tau}{max}$\\
2000 & 100 & 1 & 213 & 20 & 1& $\tau> \scrd{\tau}{max}$\\
\hline
\end{tabular}
\end{table}
\Tab \ref{tab-Q-size} illustrates the data set size $\abs{Q_k(n,\tau,s)}$
for $(N, d, s, n) = (2000, 100, 1, 213)$ and $\tau\in\set{1, 2, \cdots, 20}$. It is obvious that different time delay $\tau$ leads to different size  $\abs{Q_k(n,\tau,s)}$. Obviously, the value of $x[n]$ calculated by averaging depends on the elements of the data set $Q_k(n,\tau, s)$ and the size $\abs{Q_k(n,\tau,s)}$. In order to estimate the value of $x[n]$, it is necessary to replace the arithmetic average estimator with a better estimator, say median filter or other proper estimator. 

As the generalization of the pulling back theorem under the noisy free condition, we now give the revised version the pulling back theorem as follows:

\begin{thm} \label{thm-pullback-noisy}
For the type $s\in\set{0,1}$, embedding dimension $d\in \mathbb{N}$ and time sequence $X=\set{x[n]}^{N-1+s}_{n=s}$ of length $N$ perturbed by noise, we can pull back the $k$-th CTM $\mat{Z}_k=(Z^i_j(k))$ in the immersion space $\ES{R}{d}{m}$ such that $\displaystyle\mathcal{D}(\mat{M})=\sum^r_{k=1}\mat{Z}_k$ to the $k$-th CTS $\scrd{x}{cts}^{(k)}[n]$ by  
\begin{equation} \label{eq-xn-estimate}
\scrd{x}{cts}^{(k)}[n] = \ProcName{MeanSolver}(Q_k(n,\tau,s), \abs{Q_k(n,\tau,s)})
\end{equation}
where \ProcName{MeanSolver} is the algorithm for sequence estimation by averaging  the data set $Q_k(n,\tau,s)$.
\end{thm}   
 
Particularly, if the noisy does not exist, the filtering operator must be the identity operator and \Thm \ref{thm-pullback-noisy} degrades to \Thm \ref{thm-pullback-noisyfree} since $\mathcal{F}$ degrades to the identity operator $\mathbb{1}$ and the $\ProcName{MeanSolver}$ algorithm for state estimation can be implemented with arithmetic average algorithm $\ProcName{AriAveSolver}$ or median algorithm \ProcName{MedianSolver}. 

\subsection{Impact of Time Delay}

The time sequence $X=\set{x[n]: s\le n\le N-1+s}$ is usually obtained by sampling the continuous signal $x(t)$ with the sampling frequency $f$. Formally, we have
\begin{equation}
x[n] = x(t_0 + n/f), \quad  n\in \mathbb{Z}^+, t\in[t_0, \scrd{t}{end}] 
\end{equation}
for the initial time $t_0\in \mathbb{R}$ and final time $\scrd{t}{end}$. We have the following observation for configuring the time delay $\tau\in \mathbb{N}$:
\begin{itemize}
\item If the sampling frequency $f$ is high, then $N = \mfloor{(\scrd{t}{end}-t_0)f}$ will be large. The smallest 
$\tau=1$ means a large integer $m = N-(d-1)\tau$ for the given embedding dimension $d$ when $N$ is large, which implies the matrix $\mat{M}$ of $d$-by-$m$ is a big matrix. For the matrix decomposition involved in the mode decomposition will lead to high computational complexity. In the sense of reducing the computational complexity of mode decomposition, we should set $\tau> 1$ and a big $\tau$ may be better.
\item For the fixed length $N$, embedding dimension $d$ and time delay $\tau$, we have
\begin{equation}
\abs{Q_k(n,\tau,s)} = \scrd{q}{max}-\scrd{q}{min}+1
\end{equation}  
and for the fixed $n$ and $s$, $\abs{Q_k(n,\tau,s)}$ is decreasing when $\tau$ is increasing according to \eqref{eq-qmin-qmax}. In other words, a big $\tau$ means a small date set $Q_k(n,\tau,s)$, which lowers the performance of the estimation algorithm \ProcName{MeanSolver}. 
\item For the time sequence perturbed by noise, it is wise to set a lower bound for the number of the columns for the trajectory matrix $\mat{M}\in \ES{R}{d}{m}$, i.e.,
\begin{equation}
m \ge d \ge 2 \scrd{n}{dof} + 1,
\end{equation} 
which implies that
\begin{equation}
\tau \le \mfloor{\frac{N-d}{d-1}} \le \mfloor{\frac{N-(2\scrd{n}{dof}+1)}{d-1}}
\end{equation}
Usually, the intrinsic dimension $\scrd{n}{dof}$ is not known and we can set the upper bound as
\begin{equation}
\scrd{\tau}{max} = \mfloor{\frac{N-d}{d-1}}.
\end{equation}
As shown in \Tab \ref{tab-Q-size}, for the $(N,d) = (2000, 100)$, we have $\scrd{\tau}{max} = \mfloor{1900/99} = 19$ and $\abs{Q_k(213, 19, 1)}=7$ candidates can be for estimating the value of $x[213]$. Moreover, as shown in \Tab \ref{tab-size-mat-M}, we have $\scrd{m}{min} = 119$. Otherwise, if we take $\tau = 20$, then $m=20<d=100$ and only one candidate can be used for estimating $x[213]$, which should be avoided in statistical estimation. 
\end{itemize} 
In summary, we should balance the computational cost and the precision of the mode decomposition when configuring the time delay $\tau$ such that $1\le \tau \le \scrd{\tau}{max}$.

\subsection{Application in Singular Spectrum Analysis}

There are two significant issues that should be noted:
 firstly, the \eqref{eq-X-traj-mat-1} is widely used for the trajectory matrix in  \textit{singular spectrum analysis} (SSA); secondly, the SSA with the DAP for converting trajectory matrix to time sequence is originally proposed by Vautard  in 1992 \cite{Vautard1992SSA}. Thus we can deduce  that the SGMD and SSA share the common steps of embedding  for up conversion and the 
DAP for down conversion. For the time delay $\tau = 1$,  the SGMD and SSA have the same trajectory matrix $\mat{M} = \Phi_{s,\tau}(X)$ if the $d$ and $m$ are the same. With the help of the pulling back theorem, the embedding of SSA can be replaced with the embedding mapping $\Phi$ used in the SGMD by allowing $\tau\ge 1$.

\section{Conclusions} \label{sect-conclusions}

The original SGMD is limited to the two cases without doubts in the past five years: 
\begin{itemize}
\item time delay $\tau=1$ in the inversion of embedding step, which is due to the dependence on the diagonal averaging principle; 
\item the embedding just holds for the type-1 time sequence denoted by $X=\seq{x[1], x[2], \cdots, x[N]}$ for the Fortran/MATLAB/Octave/... programming languages and fails for the type-0 time sequence denoted by $X=\seq{x[0], x[1], \cdots, x[N-1]}$ for the C/C++/Java/Python/Rust/... programming languages.
\end{itemize}

Our main conclusions include the following significant aspects:
\begin{itemize}
\item The pulling back theorem for inverting the embedding step in SGMD is proposed for deterministic time sequences with the theory of Diophantine equation in number theory for the general case of time delay $\tau\in \mathbb{N}$ and time sequence $X=\set{x[n]}^{N-1+s}_{n=s}$ for $s\in \set{0, 1}$. 
\item In order to deal with random time sequences, the pulling back theorem is generalized by introducing a denoising step after decomposing the trajectory matrix and using a mean estimation algorithm for pulling back the CTM to the corresponding CTS. 
\item The discussion of how to configure the time delay $\tau$ in embedding step shows that small $\tau$ means better mean estimation but large computational complexity in matrix decomposition, thus a proper value $\tau$ is needed for balancing the efficiency and accuracy. 
\end{itemize}

In the future work, we will propose novel version of the algorithms for SGMD with lower computational complexity and less constraints for the time delay, and exploring the relation of SSA and SGMD with our pulling back theorem.

\subsection*{Acknowledgments}

This work was supported in part by the National Natural Science Foundation of China under grant numbers 62167003 and 62373042, in part by the Hainan Provincial Natural Science Foundation of China under grant numbers 720RC616 and 623RC480, in part by the Research Project on Education and Teaching Reform in Higher Education System of Hainan Province under grant number Hnjg2025ZD-28, in part by the Specific Research Fund of the Innovation Platform for Academicians of Hainan Province, in part by the Hainan Province Key R \& D Program Project under grant number ZDYF2021GXJS010, in part by the Guangdong Basic and Applied Basic Research Foundation under grant number 2023A1515010275, and in part by the Foundation of National Key Laboratory of Human Factors Engineering under grant number HFNKL2023WW11.

\subsection*{Data Availability}

Not applicable

\subsection*{Code Availability}

Not applicable

\subsection*{Declaration of interests}

 The authors declare that they have no known competing financial interests or personal relationships that could have appeared to influence the work reported in this paper.
 
\begin{appendix}
\section{Diophantine Equation $\tau x + y =n + s\tau $} \label{appendix-1}

For the Diophantine equation $a x + by =n$, we have the following important 
properties \cite{HuaLK1958,Cohen2007GTM239}: 

\begin{thm} \label{thm-appendix-1}
For $a, b\in \mathbb{N}=\set{1,2,3,\cdots}$ and $n\in \mathbb{Z}=\set{0, \pm 1, \pm2, \cdots}$, the equation $ax+by = n$ has solution $(x,y)\in \mathbb{Z}^2$ if and only if the greatest common divisor of $a$ and $b$ is a factor of $n$, i.e., $\GCD(a,b)\mid n$.
\end{thm}
\begin{thm} \label{thm-appendix-2}
Suppose that $a, b\in \mathbb{N}$ such that $\GCD(a,b) = 1$. For any $n\in \mathbb{Z}$ such that $n> ab -a-b$, there must exist $x, y\in \mathbb{Z}^+ = \set{0, 1, 2, \cdots}$ such that $n = ax + by$.
\end{thm}

If we take the following assignment
\begin{equation}
\left\{
\begin{aligned}
a &\gets \tau\\
b &\gets 1 \\
n &\gets n+s\tau
\end{aligned}
\right.
\end{equation}
for $s\in \set{0,1}$ and $\tau\in \mathbb{N}$, then the solution of the Diophantine equation $\tau x + y = n + s\tau$ have non-negative solution $(x,y)\in (\mathbb{Z}^+)^2$ for any $n \in \mathbb{Z}^+$. 
\end{appendix}


\end{document}